\documentclass[12pt]{article}
\usepackage{amsmath, amssymb}
\usepackage{amsfonts}
\usepackage{graphicx,psfrag,epsf}
\usepackage{enumerate}
\usepackage{natbib, xcolor}
\usepackage{url} 
\usepackage{algorithm}
\usepackage{booktabs}
\usepackage{multirow}
\usepackage{verbatim}
\usepackage{array}
\usepackage{setspace}
\usepackage{authblk}
\usepackage[compact]{titlesec}
\usepackage[nomarkers, nolists]{endfloat}
\usepackage{lscape} 
\usepackage{rotating}

\usepackage{booktabs} 
\usepackage{tabularx} 
\usepackage{caption}  

\newcommand{\blind}{0}

\newcolumntype{L}[1]{>{\raggedright\let\newline\\\arraybackslash\hspace{0pt}}m{#1}}
\newcolumntype{C}[1]{>{\centering\let\newline\\\arraybackslash\hspace{0pt}}m{#1}}
\newcolumntype{R}[1]{>{\raggedleft\let\newline\\\arraybackslash\hspace{0pt}}m{#1}}

\begin{document}

\def\spacingset#1{\renewcommand{\baselinestretch}%
{#1}\small\normalsize}
\spacingset{1}


\if0\blind
{

\title{\bf A sensitivity analysis for non-inferiority studies with non-randomised data}
\author[1*]{Daijiro Kabata, PhD, MPH}
\author[2*]{Takumi Imai, PhD}

\affil[1]{\small Center for Mathematical and Data Science, Kobe University, Kobe, Hyogo, Japan}
\affil[2]{\small Clinical \& Translational Research Center, Kobe University Hospital, Kobe, Hyogo, Japan}
\affil[*]{{\small The authors are co‑corresponding authors with equal contribution.

\textit{email:} daijiro.kabata@port.kobe-u.ac.jp}}

\date{}
  \maketitle
} \fi

\bigskip

\begin{abstract}
\noindent
\textbf{Background}
Non-inferiority studies based on non-randomised data are increasingly used in clinical research but remain prone to unmeasured confounding. The classical E-value offers a simple way to quantify such bias but has been applied almost exclusively with respect to the statistical null. We reformulated the E-value framework to make explicit its applicability to predefined clinical margins, thereby extending its utility to non-inferiority analyses.

\noindent
\textbf{Development}
Using the bias-factor formulation by Ding and VanderWeele, we defined the non-inferiority E-value as the minimum strength of association that an unmeasured confounder would need with both treatment and outcome, on the risk-ratio scale, to move the 95\% confidence-limit estimate to the prespecified non-inferiority margin. 

\noindent
\textbf{Application}
This approach was applied to three observational studies and one single-arm trial with external controls to illustrate interpretation and range. The resulting non-inferiority E-values for the confidence limits varied from about one to three, depending on design and findings. In the single-arm trial, a large gap between the confidence-limit and point-estimate NIEs reflected small sample size and wide confidence intervals, highlighting that both should be reported for a balanced assessment of robustness.

\noindent
\textbf{Conclusion}
This study reformulates the E-value to focus on clinically meaningful margins rather than the statistical null, enabling its application to non-inferiority analyses. Although the non-inferiority E-value inherits the limitations of the original method and cannot address all bias sources, it offers a transparent framework for interpreting non-randomised evidence and for generating insights that inform the design of future, more definitive randomised controlled trials.

\end{abstract}

\noindent%
\vfill

\newpage
\spacingset{1.45}

\section{Introduction}
\label{sec:intro}

Non-randomised studies increasingly inform comparative effectiveness and safety research, particularly when randomised trials are infeasible, unethical, or require very large sample sizes. Such designs—including observational studies and single-arm trials with external controls—are, however, vulnerable to unmeasured confounding, which conventional regression or propensity-score methods cannot address. In the absence of a valid instrumental variable, sensitivity analyses are therefore essential to assess how unmeasured confounding might alter causal conclusions.

For assessing the impact of unmeasured confounders, \cite{VanderWeele2017-jj} introduced the value of E‑value as the minimum strength of association that an unmeasured confounder would need with both the exposure and the outcome to explain the absence of an observed effect as no effect. 
Larger E-values indicate that only a strong unmeasured confounder could plausibly shift the observed association to the null.
Despite critiques that it rests on strong assumptions and is essentially a transformation of the effect estimate that does not address other sources of bias, it has been widely adopted in observational epidemiology as a simple, standardised summary of robustness to unmeasured confounding.\citep{Blum2020-qp,Ioannidis2019-fv,Greenland2020-yd,Poole2020-qk} Fox and colleagues further argued that most applications focus on sensitivity to the statistical null hypothesis (\(M=1\)), whereas clinical decisions often hinge on whether effects exceed or remain within a prespecified margin.\citep{Fox2020-vp} This highlights the need to adapt the E‑value framework to clinically defined margins rather than only the null, such as the non‑inferiority margins considered in this paper.

Non-inferiority trials differ from conventional superiority trials in both scientific aim and interpretation. In a superiority trial the primary question is whether the new treatment is better than the control, and inference typically focuses on whether the confidence interval excludes the null value (for example, a risk ratio of 1). In contrast, non-inferiority trials are conducted when an effective standard treatment already exists, and the goal is to show that a new treatment is not unacceptably worse, often because it may offer advantages in safety, convenience, or cost. This is formalised through a non-inferiority margin $M$, chosen to represent the largest loss of effectiveness that would still be clinically acceptable. The primary analysis asks whether the confidence interval for the treatment effect lies entirely within the region compatible with non-inferiority (for example, below $M$ for adverse outcomes or above $M$ for beneficial outcomes), rather than whether it excludes the null.

This shift from the null value to a clinically defined margin has important implications for sensitivity analyses. In non-inferiority studies, the key question is whether unmeasured confounding could move the effect estimate or its governing confidence bound across the prespecified margin $M$, thereby changing the clinical interpretation from non-inferior to not yet established as non-inferior. Using the conventional E-value in this context therefore targets a different question from the one underlying the non-inferiority decision. To align sensitivity analyses with that decision problem, we adapt the E-value framework so that it quantifies how strong an unmeasured confounder would need to be to move the relevant confidence limit to the non-inferiority margin. We refer to this quantity as the non-inferiority E-value (NIE).

Non-inferiority designs are often adopted in disease areas where standard treatments are well established, particularly when new interventions offer clear advantages over standard care, such as reduced dosing frequency, improved safety, lower cost, or greater convenience \citep{Mauri2017-jl}. The large sample sizes often required for non-inferiority trials, especially with binary or survival endpoints, have motivated greater use of non-randomised designs, including observational studies and single-arm trials with external controls. For such studies, transparent assessment of vulnerability to unmeasured confounding is critical.

In this article, we revisit the bias-factor formulation of \cite{Ding2016-uo} and derive the NIE as the minimum strength of unmeasured confounding needed to move a confidence-limit estimate to the prespecified non-inferiority margin. We then apply this framework to published non-randomised non-inferiority studies to illustrate computation and interpretation. Our aim is to provide investigators with a simple tool to assess how unmeasured confounding could overturn or support non-inferiority conclusions in non-randomised settings.

\newpage
\section{E-value framework revisited}
\label{sec:methods}

\subsection{Brief Review of E-value}

\begin{figure}[h]

        \caption{Variable relationship and components of the non-inferiority E-value}
        \centering
        \includegraphics[width=\textwidth]{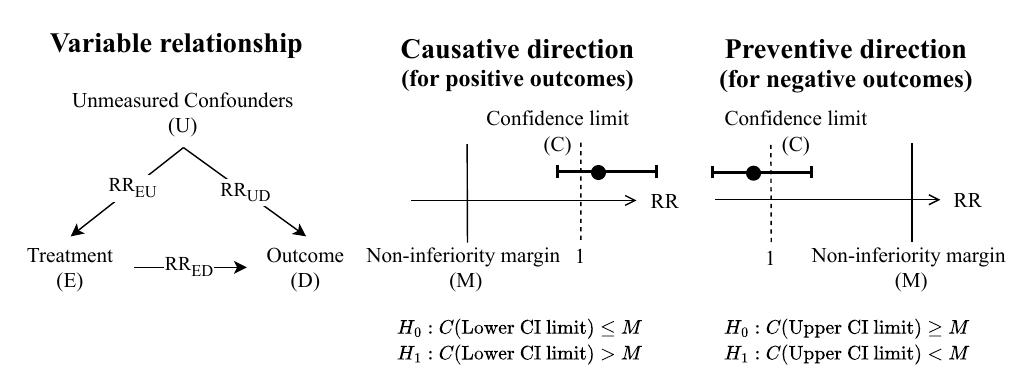}
        
        \vspace{15pt}
        
        \raggedright
        Directed acyclic graph (left) depicts an unmeasured confounder \(U\) associated with both exposure \(E\) and outcome \(D\).
        The bias factor for multiplicative confounding is given by $B = \frac{RR_{UD} \times RR_{EU}}{RR_{UD} + RR_{EU} - 1}$. 
        The NIE quantifies the minimum strength of association an unmeasured confounder must have with both \(E\) and \(D\) to move the governing 95\% confidence-limit estimate \(C\) to the prespecified non-inferiority margin \(M\): $\mathrm{NIE}(C, M) = \kappa + \sqrt{\kappa(\kappa - 1)}$ where $\kappa = \max\left(\frac{C}{M}, \frac{M}{C}\right) \ge 1$.
        For causative-direction hypotheses (center), \(C\) is the lower 95\% CI bound and $\kappa = \frac{C}{M}$; for preventive-direction hypotheses (rihgt), \(C\) is the upper CI bound and $\kappa = \frac{M}{C}$.
        \newline
        Abbreviations: RR, risk ratio; CI, confidence interval; NIE, non-inferiority E-value.
        
        \label{fig:dag}
\end{figure}

This section provides a brief review of the seminal paper by \cite{Ding2016-uo} and \cite{VanderWeele2017-jj}, and discusses key aspects essential for understanding the application of the E-value in non-inferiority assessment.
The basic notation follows \cite{Ding2016-uo}: let E denote the exposure, D denote a binary outcome, 
and U denote unmeasured confounders. The two association parameters, $\mathrm{RR}_{EU}$ and $\mathrm{RR}_{UD}$, 
characterizing the relationships between E and U, and between U and D, respectively (Figure\ref{fig:dag}). They are also adopted as defined in \cite{Ding2016-uo}.

Prior to discussing the E-value, we briefly highlight an important implication concerning 
the bounding factor (bias factor) defined below.
\begin{equation}
  B \equiv \frac{\mathrm{RR}_{UD} \times \mathrm{RR}_{EU}}{\mathrm{RR}_{UD} + \mathrm{RR}_{EU} - 1} \nonumber
\end{equation}
Under multiplicative confounding, the maximal amount of bias induced by U can be bounded by this factor; 
assuming a causative direction, the ratio of $\mathrm{RR}_{ED}^\mathrm{obs}$ to $\mathrm{RR}_{ED}^\mathrm{true}$ is 
bounded from above by B (Equation 1 in \cite{Ding2016-uo}), whereas assuming a preventive direction, 
the ratio is bounded from below by 1/B (Equation 2 in \cite{Ding2016-uo}).
\begin{align}
  \mathrm{RR}_{ED}^\mathrm{obs} / \mathrm{RR}_{ED}^\mathrm{true} &\le B  && (\text{assuming a causative direction}) 
  \label{eq:bounded1} \\
  \mathrm{RR}_{ED}^\mathrm{obs} / \mathrm{RR}_{ED}^\mathrm{true} &\ge 1/B && (\text{assuming a preventive direction})
  \label{eq:bounded2}
\end{align}
Assuming $\mathrm{RR}_{ED}^\mathrm{true}=1$, the expressions further simplify: for a causative direction, $\mathrm{RR}_{ED}^\mathrm{obs} \le B$; whereas for a preventive direction, $\mathrm{RR}_{ED}^\mathrm{obs} \ge 1/B$. 
Given an observed $\mathrm{RR}_{ED}^\mathrm{obs}$, the region of ($\mathrm{RR}_{EU}$, $\mathrm{RR}_{UD}$) 
satisfying these inequalities can be interpreted as representing the characteristics of an unmeasured confounder that 
could produce the observed $\mathrm{RR}_{ED}^\mathrm{obs}$ under the condition that $\mathrm{RR}_{ED}^\mathrm{true}=1$. 
In other words, if it is probable that an unmeasured confounder lies within the region of ($\mathrm{RR}_{EU}$, 
$\mathrm{RR}_{UD}$) satisfying these inequalities, the observed risk ratio $\mathrm{RR}_{ED}^\mathrm{obs}$ could be completely spurious.
nmnThe E-value provides a simplified approach to such an assessment by indicating the minimum value of 
max($\mathrm{RR}_{EU}$, $\mathrm{RR}_{UD}$) needed to shift the observed $\mathrm{RR}_{ED}^\mathrm{obs}$ 
(point estimate or confidence limit) to the null. The minimum value of 
max($\mathrm{RR}_{EU}$, $\mathrm{RR}_{UD}$) is achieved when $\mathrm{RR}_{EU}=\mathrm{RR}_{UD}$ 
at the boundary of the inequalities described above, and solving $\mathrm{RR}_{ED}^\mathrm{obs} = B$ or 
$\mathrm{RR}_{ED}^\mathrm{obs} = 1/B$ under this condition yields the following expression for the classical E-value.
\begin{align}
  \textrm{E-value} &= \mathrm{RR}_{ED}^\mathrm{obs} + \sqrt{\mathrm{RR}_{ED}^\mathrm{obs} \times (\mathrm{RR}_{ED}^\mathrm{obs} - 1)}  && (\text{assuming a causative direction}) 
  \label{eq:classicalEvalue1} \\
  \textrm{E-value} &= 1/\mathrm{RR}_{ED}^\mathrm{obs} + \sqrt{1/\mathrm{RR}_{ED}^\mathrm{obs} \times (1/\mathrm{RR}_{ED}^\mathrm{obs} - 1)} && (\text{assuming a preventive direction})
  \label{eq:classicalEvalue2}
\end{align}
The interpretation of the E-value with illustrative examples is presented in \cite{VanderWeele2017-jj}.

Calculations for effect measures other than the risk ratio (RR) have also been discussed and are summarized 
in Table 2 of \cite{VanderWeele2017-jj}. For instance, with respect to the hazard ratio (HR)—which will also be examined 
in the subsequent case studies—it has been suggested that when the outcome is relatively rare 
(e.g., occurring in less than $15\%$ of participants by the end of follow-up), the aforementioned E-value formula 
for the RR can be applied directly. In contrast, when the outcome is more common (e.g., $\ge 15\%$ by the end of follow-up), 
it is recommended to first convert the HR to an approximate RR using the following equation, 
and then compute the corresponding E-value.
\begin{equation}
\mathrm{RR} \approx \frac{1 - 0.5^{\,\sqrt{\mathrm{HR}}}}{1 - 0.5^{\,\sqrt{1/\mathrm{HR}}}}
\end{equation}

\subsection{E‑value for non-inferiority designs}

In non-inferiority analysis we consider the risk ratio as $\mathrm{RR}_{ED}=P(D=1|E=1)/P(D=1|E=0)$. 
The hypothesis orientation is defined relative to the outcome as coded. 
We use causative direction when letting $D=1$ for clinical success, and preventive direction when letting $D=1$ for an adverse event. 
For a causative-direction hypothesis in non-inferiority analysis, we use a typical margin $M\le1$ and test $H_0:\mathrm{RR}_{ED}^{true}\le M$ versus $H_1:\mathrm{RR}_{ED}^{true}>M$ and compare the lower 95\% confidence limit with $M$. 
For a preventive-direction hypothesis, we use a typical margin $M\ge1$ we test $H_0:\mathrm{RR}_{ED}^{true}\ge M$ versus $H_1:\mathrm{RR}_{ED}^{true}<M$ and compare the upper 95\% confidence limit with $M$ (see Figure\ref{fig:dag}). 
If the outcome coding is reversed, the labels “causative” and “preventive” are interchanged, but the rule—comparing the governing CI limit with $M$--and the definition of the non‑inferiority E‑value are unchanged. Thus, $M$ serves as the reference point regardless of coding.
Therefore, the conventional E-value derived under the assumption of $\mathrm{RR}_{ED}^{\mathrm{true}} = 1$ is not 
appropriate in the context of non-inferiority assessment, and it is necessary to revisit the definition of 
the E-value based on Equations \eqref{eq:bounded1} and \eqref{eq:bounded2}. Without assuming $\mathrm{RR}_{ED}^{\mathrm{true}} = 1$, if we seek the minimum value of $\max(\mathrm{RR}_{EU}, \mathrm{RR}_{UD})$ needed to shift the observed $\mathrm{RR}_{ED}^{\mathrm{obs}}$ to the $\mathrm{RR}_{ED}^{\mathrm{true}}$, the following definition is obtained (a similar mathematical formulation can also be found in Equation~6 of Ding and VanderWeele (2016)).
\begin{align}
  \textrm{E-value} &= \Big\{ \mathrm{RR}_{ED}^\mathrm{obs} + \sqrt{\mathrm{RR}_{ED}^\mathrm{obs} \times (\mathrm{RR}_{ED}^\mathrm{obs} - \mathrm{RR}_{ED}^\mathrm{true})} \Big\} \Big/ \mathrm{RR}_{ED}^\mathrm{true} && (\text{assuming a causative direction}) \\
  \textrm{E-value} &= \Big\{ 1/\mathrm{RR}_{ED}^\mathrm{obs} + \sqrt{1/\mathrm{RR}_{ED}^\mathrm{obs} \times (1/\mathrm{RR}_{ED}^\mathrm{obs} - 1/\mathrm{RR}_{ED}^\mathrm{true})} \Big\} \times \mathrm{RR}_{ED}^\mathrm{true} && (\text{assuming a preventive direction})
\end{align}
The E-value in the context of non-inferiority can be calculated by substituting 
the pre-specified non-inferiority margin for $\mathrm{RR}_{ED}^{\mathrm{true}}$ and 
the 95\% confidence limit of the effect measure used for comparison with the non-inferiority margin for $\mathrm{RR}_{ED}^{\mathrm{obs}}$.  
This provides the minimum value of $\max(\mathrm{RR}_{EU}, \mathrm{RR}_{UD})$ that could generate the observed $\mathrm{RR}_{ED}^{\mathrm{obs}}$ under the assumption that the  represents the true effect, which is the E-value for non-inferiority.

Prior to the case studies, a further notational simplification is applied.  
The 95\% confidence limit of the effect measure being compared to the non-inferiority margin is denoted by $C$.  
For notational convenience, the following measure is defined to evaluate the distance between the non-inferiority margin and the 95\% confidence limit on a multiplicative scale.
\begin{equation}
  \kappa = \max\!\left(\frac{C}{M}, \frac{M}{C}\right),
\end{equation}
which satisfies $\kappa \geq 1$. 
The quantity $\kappa$ measures the relative difference between $C$ and $M$: 
when $C > M$, it indicates how much greater $C$ is than $M$, and when $C < M$, how much less $C$ is than $M$.  
For a causative direction, non-inferiority conclusions correspond to $C > M$ (the lower 95\% CI bound is above M), so $\kappa = C/M$; 
for a preventive direction, non-inferiority conclusions correspond to $C < M$ (the upper 95\% CI bound is below M), so $\kappa = M/C$.
The NIE can be expressed as follows.
\begin{equation}
  \mathrm{NIE}(C, M) = \kappa + \sqrt{\kappa\,(\kappa - 1)}.
  \label{eq:nev}
\end{equation}
Because the NIE depends only on $\kappa=max(C/M, M/C)$, it is invariant to how the binary outcome is coded; only the governing CI limit (upper vs lower) changes with the effect direction.
When $M = 1$ this reduces to the classical E‑value \eqref{eq:classicalEvalue1}\eqref{eq:classicalEvalue2}.  Equation\,\eqref{eq:nev}
gives the minimum magnitude that $\max(\mathrm{RR}_{UD}, \mathrm{RR}_{EU})$ must
attain to move the confidence limit to equal $M$.  
The NIE can also be computed for the point estimate of the effect measure to quantify the robustness of the
estimated effect itself.

Similarly, given $M$ and $C$, the region of $(\mathrm{RR}_{EU}, \mathrm{RR}_{UD})$ satisfying Equations (1) and (2) 
can be obtained using the following simplified expressions.
\begin{equation}
  \kappa = \frac{\mathrm{RR}_{UD}\,\mathrm{RR}_{EU}}{\mathrm{RR}_{UD} + \mathrm{RR}_{EU} - 1}.
  \label{eq:biasfactorNIE}
\end{equation}
This contour forms a hyperbola-like curve in the $(\mathrm{RR}_{UD}, \mathrm{RR}_{EU})$ plane.  
Values of $(\mathrm{RR}_{UD}, \mathrm{RR}_{EU})$ on or above the contour represent combinations of 
confounding strengths sufficient to shift the confidence limit to $M$, 
whereas values below the contour cannot fully account for the non-inferiority conclusion.  
These curves are illustrated in our case studies.

\newpage
\section{Case studies}
\label{sec:examples}

To demonstrate our proposed methodology we focused on three observational non-inferiority studies published between 2023 and 2025.  
These studies were identified
through literature searches of major medical journals and were selected
because they reported adjusted HRs with two‑sided 95\% confidence
intervals and pre‑specified non-inferiority margins.  
The selected investigations include an analysis of kidney transplantation from donors with and without HIV, a comparison of hybrid versus flat dosing of pembrolizumab for
non‑small cell lung cancer, and an evaluation of delayed versus timely
curative treatment for colon cancer.  Each study provided sufficient
information to compute NIEs, namely the adjusted effect estimate (hazard
ratio), the corresponding confidence interval limits and the non-inferiority margin. 
Table\ref{tab:summary} presents the main characteristics and results of the selected studies. 
All four case studies are analyzed under the preventive direction; hence we compare the upper 95\% CI limit with $M$ and use $\kappa=M/C$ throughout. 

\subsection{\cite{Durand2024-ta}; Kidney transplantation from donors with and without HIV}

\begin{figure}[h]

        \caption{Bias-factor contours and NIE values across four non-randomised non-inferiority case studies}
        \centering
        \includegraphics[width=\textwidth]{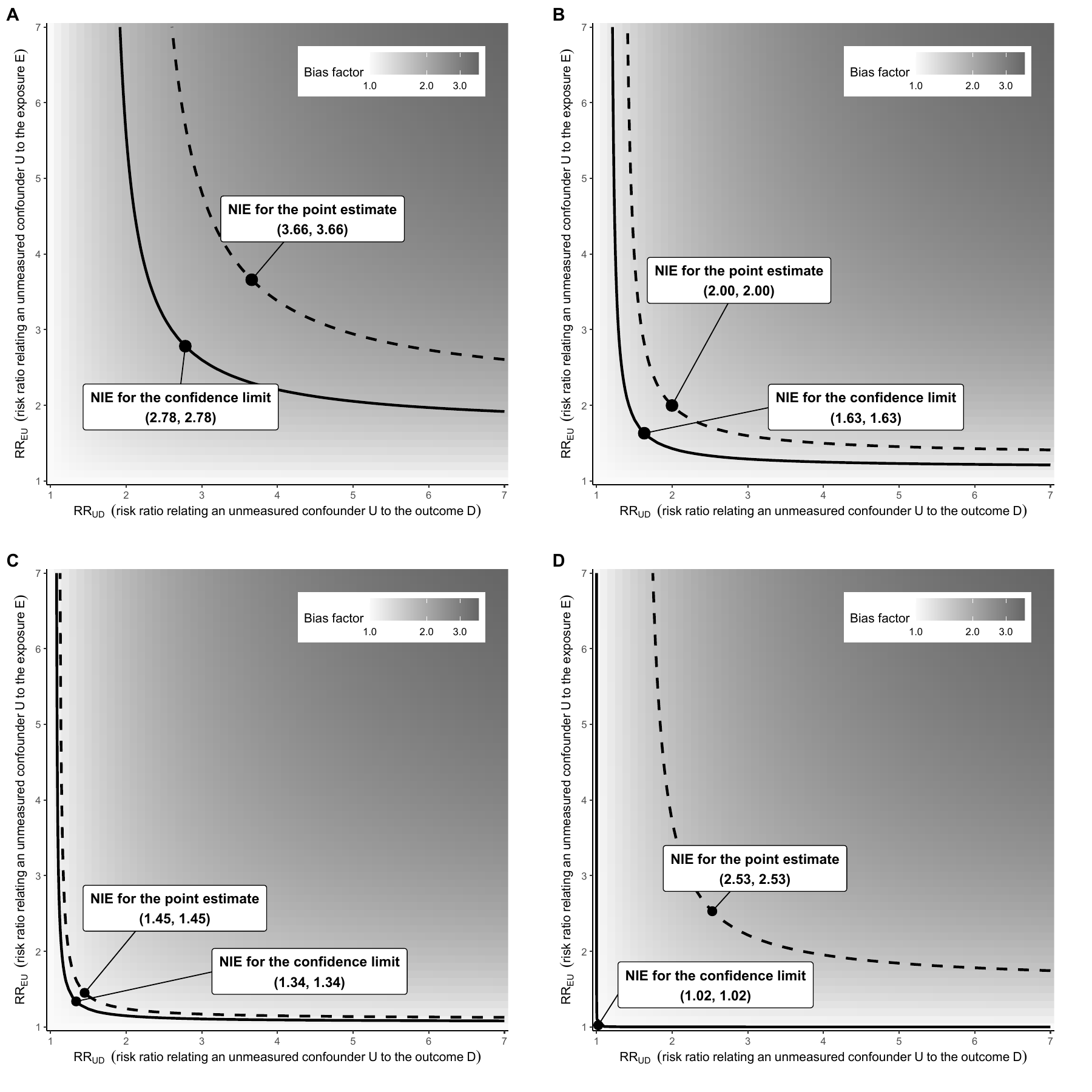}
        
        \vspace{15pt}
        \raggedright
        Panels A–D plot \(RR_{EU}\) (x-axis) against \(RR_{UD}\) (y-axis). 
        Shading indicates the bias factor \(B\).
        The solid curve shows combinations of \((RR_{EU}, RR_{UD})\) sufficient to move the governing 95\% CI limit to the non-inferiority margin \(M\); the dashed curve does so for the point estimate. 
        The labelled markers denote the corresponding NIE values where \(RR_{EU} = RR_{UD} = \mathrm{NIE}\). 
        (A) Kidney transplantation from donors with and without HIV; NIEs 2.78 (CI limit) and 3.66 (point). (B) Hybrid vs flat dosing of pembrolizumab in advanced non–small cell lung cancer; NIEs 1.63 and 2.00. (C) Delayed vs timely curative treatment for colon cancer; NIEs 1.34 and 1.45. (D) Robotic vs laparoscopic total gastrectomy using an external control cohort; NIEs 1.02 and 2.53.
        \newline
        Abbreviations: RR, risk ratio; HR, hazard ratio; CI, confidence interval; NIE, non-inferiority E-value. 
        
        \label{fig:bf}
\end{figure}

\cite{Durand2024-ta} evaluated the safety of transplanting kidneys from
HIV‑positive donors into HIV‑positive recipients.  In their cohort of 99
recipients of HIV‑positive donor kidneys and 99 recipients of HIV‑negative
kidneys, the primary composite outcome comprised death, graft loss, serious
adverse events, HIV breakthrough infection, persistent virologic failure or opportunistic infection.  The adjusted HR was 1.00 with a 95\% CI 0.73--1.38 and a pre‑specified non-inferiority margin of 3.0
\citep{Durand2024-ta}.  Because the composite outcome was relatively frequent (79/99 and 77/99 among recipients with HIV-positive and -negative kidneys, respectively), we approximated the HR as an RR using the formula described in Section\,\ref{sec:methods}.  The original upper CI bound of 1.38 corresponds to an approximate RR of 1.25.  Furthermore, the non-inferiority margin of 3.0 on the HR was converted to 2.12 on the RR. This gives $\kappa = 2.12/1.25 \approx 1.7$ and an NIE of 2.78 for the confidence limit.  Similarly, the point estimate HR of 1.00 yields an approximate RR of 1.00 and $\kappa = 2.12/1.00$, giving an NIE of 3.66 (Figure\ref{fig:bf}A).  These values imply that only very strong unmeasured confounding—on the order of 2.78‑ to 3.66‑fold associations with both donor type and the composite outcome—could shift the result to the margin.  In the context of kidney transplantation, unmeasured factors such as socioeconomic status or co‑infection are unlikely
to exert such strong associations after adjustment for measured variables,
suggesting that the non‑inferiority conclusion is robust.

\subsection{\cite{Smeenk2025-nk}; Hybrid and flat dosing of pembrolizumab}

\cite{Smeenk2025-nk} examined a hybrid weight‑based dosing regimen of
pembrolizumab compared with the standard flat dose in a retrospective cohort
of advanced non‑small cell lung cancer patients.  Their inverse probability
of treatment weighted analysis yielded an adjusted HR of 0.76
(95\% CI 0.63--0.91) with a non-inferiority margin of 1.15
\citep{Smeenk2025-nk}.  The outcome of overall survival is common in this
population (256/375 and 320/391 among the hybrid‑dose and flat‑dose cohorts, respectively), so we applied the HR–to–RR approximation.
The upper confidence limit HR of 0.91 corresponds to an approximate risk
ratio of 0.94. Comparing this with the approximated RR margin of 1.1 yields $\kappa =
1.1/0.94 \approx 1.17$ and an NIE of 1.63.  For the point estimate HR of
0.76 the approximate RR is 0.83; comparing this with the margin
gives $\kappa = 1.1/0.83 \approx 1.33$ and an NIE of 2.0 (Figure\ref{fig:bf}B).  These values
are smaller than those observed in the kidney transplant study.  A
confounder associated with both the probability of receiving hybrid dosing
and survival by roughly 1.33‑ to 2.0‑fold could shift the CI to include the
margin or attenuate the point estimate to the margin.  Factors such as
disease severity, comorbidities or clinician preference may influence both
treatment choice and outcome and are not perfectly measured.  Consequently,
while the non‑inferiority conclusion appears reasonable, it is somewhat sensitive to
unmeasured confounding and should be interpreted cautiously.

\subsection{\cite{Rydbeck2023-ao}; Delays to curative treatment for colon cancer}

\cite{Rydbeck2023-ao} investigated whether prolonged waiting time before
curative therapy influences survival among patients with colon
cancer \citep{Rydbeck2023-ao}.  They compared patients starting treatment within 28-56 days after
diagnosis with those starting within 28 days and set a non-inferiority margin of 1.10.
The adjusted HR was 0.95 with 95\% CI 0.89--1.00, and the
investigators concluded that they could establish non‑inferiority.
Applying the HR–to–RR approximation, the upper CI bound
of 1.0 corresponds to an approximate RR of 1.0.  Comparing this
with the approximated RR margin of 1.07 gives $\kappa = 1.07/1.0 \approx 1.07$ and an NIE of 1.34.
For the point estimate HR of 0.95 the approximate RR is 0.97, and
$\kappa = 1.07/0.97 \approx 1.11$, yielding an NIE of 1.45 (Figure\ref{fig:bf}C).  These NIEs
near unity indicate that very modest unmeasured confounding—associations on
the order of 1.34‑ to 1.45-fold could move the effect estimate or CI to include the margin.  
Consequently, non-inferiority was thought to be fragile and a relatively small confounder could plausibly change the conclusion.

\subsection{\cite{Zhong2024-ty}; Robotic total gastrectomy versus laparoscopic total gastrectomy}

\cite{Zhong2024-ty} conducted a prospective single-arm trial with external controls that compared prospectively collected data from patients undergoing robotic total gastrectomy (RTG) with a historical control arm of patients who underwent laparoscopic total gastrectomy (LTG) \citep{Zhong2024-ty}. 
The control patients were extracted from the whole control cohort based on the propensity score. Specifically, the LTG group, which had similar propensity scores to the RTG group, was extracted and matched at a 1:2 ratio (RTG $n = 48$, LTG $n = 96$).
Among the matched cohort, the HR was 0.69 with 95\% CI 0.36--1.329. The non-inferiority margin was 1.33, thereby establishing non‑inferiority. 
Because the outcome frequency exceeded 15\%, we applied the HR‑to‑RR approximation. 
The upper CI bound HR 1.329 corresponds to $RR \approx 1.217$; the margin HR 1.33 corresponds to $RR = 1.218$. 
Hence $\kappa = 1.218/1.217 \approx 1.0$ and the NIE for the confidence limit was 1.02 (Figure\ref{fig:bf}D). 
The point estimate HR 0.69 corresponds to $RR \approx 0.77$; $\kappa = 1.218/0.77 \approx 1.58$ gives a point-estimate NIE of 2.53. 
These results imply that even a very weak unmeasured confounder could overturn the non-inferiority conclusion at the governing confidence bound, whereas the relatively large NIE for the point estimate suggests that its validity could be more readily substantiated in studies with larger sample sizes.

\newpage
\section{Discussion}
\label{sec:discussion}

We reformulated the E-value to address non-inferiority questions, defining the NIE as the minimum strength an unmeasured confounder must have to move the confidence limit to the margin $M$.
Applying this framework to four non-randomised studies demonstrated how the NIE quantifies diverse levels of robustness. The kidney transplant study \citep{Durand2024-ta} exhibited high robustness (confidence limit NIE $\approx$ 2.8), whereas the colon cancer study \citep{Rydbeck2023-ao} proved fragile (NIE $\approx$ 1.3), confirming vulnerability to modest confounding. The pembrolizumab study \citep{Smeenk2025-nk} fell in between, indicating moderate sensitivity.

Furthermore, in the robotic total gastrectomy study (\cite{Zhong2024-ty}), the NIE for the confidence limit was approximately 1.0, suggesting the non-inferiority conclusion could be overturned by very weak unmeasured confounding, whereas the point estimate NIE was approximately 2.5. Such a "divergence" between a small confidence limit NIE and a large point estimate NIE is likely due primarily to sample size or data variability. That is, if the sample size is small (or variability is large), the confidence interval becomes wide, causing the confidence limit used for the non-inferiority judgment to approach the margin $M$. As a result, the confidence limit NIE approaches around 1.0 (appearing fragile). This divergence suggests that a larger validation study, by reducing the confidence interval width, might allow the confidence limit NIE to approach the relative robustness suggested by the point estimate NIE; this indicates that increasing the sample size could enhance the robustness of the conclusion. Therefore, it is desirable to report both the "governing confidence limit NIE," which is directly linked to the non-inferiority decision, and the "point estimate NIE," which informs the substantive interpretation of the effect size and suggests designs for future studies. 
This discussion is consistent with previous work emphasizing the importance of evaluating the E-value not only for the point estimate but also for the confidence interval limit \cite{Trinquart2019-hx}. Specifically, calculating the E-value based on the confidence limit allows a more conservative and design-sensitive assessment of robustness to unmeasured confounding by better reflecting the influence of random variation and sample size.

From a practical standpoint for non-inferiority, while the entire confidence interval must first fall within the acceptance region, a larger NIE indicates greater resilience of the conclusion to unmeasured confounding. \cite{Blum2020-qp} reported that while associations of 1.5--2 are often observed in many applied fields, stronger associations exceeding 3 are rare after adjusting for measured confounders. Therefore, following their report, an NIE around 2 cannot automatically be called "robust," whereas an NIE exceeding 3 suggests that a practically strong confounder would have to be assumed to overturn the conclusion. 
However, as they and \cite{Fox2020-vp} point out, it is inappropriate to interpret the E-value solely by its numerical value. Only by comparing it with the potential impact of known confounders can it provide information about factors that should be adjusted for in future research. Critical discussions, such as by \cite{Ioannidis2019-fv}, also raise concerns that the E-value is merely a transformation of the observed effect size, and that its automatic calculation could become an excuse to avoid deeper thinking about confounding (e.g., \cite{Poole2020-qk}). As \cite{Fox2020-vp} state, the E-value is useful for concisely indicating sensitivity to unmeasured confounding, but it does not explain the complete bias picture. In particular, it does not account for other sources of error (e.g., selection bias, measurement error, time-dependent confounding, or model misspecification), as noted by \cite{Lash2021-ri}, making a comprehensive assessment including these factors essential. Furthermore, E-values and NIEs assume multiplicative bias from a single or composite confounder \citep{Greenland2020-yd} and provide limited information when multiple factors act simultaneously. Therefore, these metrics are not confirmatory. Overly definitive interpretations using E-values should be avoided.

A small NIE in a non-inferiority study does not necessarily imply low study quality. In a study with minimal unmeasured confounding, designed with the minimum necessary sample size, the confidence limit may be close to the margin, resulting in a small NIE. In this case, the small NIE indicates small statistical headroom, but it could also reflect an efficient design.
Conversely, the margin must be predefined based on past treatment effects and clinical justification \citep{Mauri2017-jl, Althunian2017-dv}; an overly lax margin could inflate the NIE and create an artificial appearance of robustness. 
Meanwhile, the importance of margin setting is highlighted from another perspective. \cite{Fox2020-vp} raised a significant concern about the classical E-value: that it focuses excessively on sensitivity relative to the "statistical null hypothesis" ($M=1$). In clinical decision-making, the "clinically acceptable difference (margin $M$)" is often more important than the "absence of a statistical difference ($M=1$)," yet the traditional E-value did not directly address this practical criterion. In contrast, non-inferiority testing is inherently a framework based on a margin $M$ defined as a "clinically meaningful difference." The NIE proposed in this study directly evaluates robustness against this clinically defined $M$. Therefore, when $M$ is appropriately set based on clinical justification, using the NIE helps shift the focus of sensitivity analysis from "statistical significance ($M=1$)" to "clinical acceptability ($M$)." This may mitigate the "fixation on the statistical null" concern raised by \cite{Fox2020-vp} regarding the conventional E-value.

Although the NIE is a single summary metric with strong limitations, it provides concrete information for inspecting the validity of results and for planning future studies. In particular, evaluating the robustness of non-inferiority studies (e.g., observational studies or a single-arm trial with external controls) conducted due to resource constraints or other difficulties with randomization can provide important preliminary information for planning a future RCT. For example, assessing the divergence between the point estimate NIE and the confidence limit NIE derived from the current study can help distinguish whether the result is attributable to unmeasured confounding or merely statistical precision (sample size). This, in turn, can contribute to the setting of a clinically appropriate margin and the sample size design for a future RCT aimed at achieving a robust conclusion. 
While we do not excessively advocate for the E-value, from the perspective of accumulating evidence, we believe a transparent metric like the NIE has practical value---it is "better to have it than not." When using NIE in practice, it is crucial to ensure design validity and combine it with other sensitivity analyses. It is desirable to improve balance at the design stage using some cousal-inference approaches and check balance of the important covariates. When using real-world data or external data as controls, it is necessary to verify alignment with the target population using transportability or generalisability weighting \citep{Westreich2017-rp}. Furthermore, complementing this with negative controls \citep{Lipsitch2010-or} or quantitative bias analysis enables a comprehensive judgment that does not rely excessively on the NIE value alone.

\section{Conclusion}
\label{sec:conclusion}

This study proposes a new method for quantifying the impact of unmeasured confounding in non-inferiority analyses using a scale aligned with the clinical margin. The NIE is a useful metric for concisely and intuitively indicating the robustness of non-inferiority conclusions in non-randomised studies. By reporting both the confidence limit and point estimate NIEs, explicitly justifying the margin, and combining design-stage confounding control with complementary sensitivity analyses, the transparency and reproducibility of non-inferiority research can be enhanced.


\newpage
\section*{Acknowledgements}

We thank the authors of the original studies for making their results
publicly available. This work involved no access to individual patient data.

\section*{Fundings}
This work is supported by JSPS KAKENHI (23K17245).

\newpage
\section*{Tables}

\begin{sidewaystable}[htbp]
\centering
\caption{Summary of Non-inferiority Studies}
\label{tab:summary}
\begin{tabular}{
  >{\raggedright\arraybackslash}p{3cm}
  >{\raggedright\arraybackslash}p{4.5cm} 
  >{\raggedright\arraybackslash}p{4.5cm} 
  >{\centering\arraybackslash}p{3.5cm} 
  >{\centering\arraybackslash}p{2.5cm}   
  >{\centering\arraybackslash}p{2.5cm}   
}

\toprule
\textbf{Study} & 
\textbf{Population; outcome} & 
\textbf{Exposure vs. Comparator} & 
\textbf{Estimated HR (95\% CI); a non-inferiority margin} & 
\textbf{NIE for the confidence limit} & 
\textbf{NIE for the point estimate} \\
\midrule
Durand \emph{et al.}\ (2024) & 
HIV-positive kidney transplant recipients; Composite safety event &
HIV-positive donor kidney vs. HIV-negative donor kidney &
1.00 (0.73, 1.38); 3.0 &
2.78 &
3.66 \\
\addlinespace
Smeenk \emph{et al.}\ (2025) &
Advanced non-small cell lung cancer patients; All-cause mortality &
Hybrid dose of pembrolizumab vs. flat dose of pembrolizumab &
0.76 (0.63, 0.91); 1.15 &
1.63 &
2 \\
\addlinespace
Rydbeck \emph{et al.}\ (2023) &
Colon cancer patients; All-cause mortality &
Treatment start within 29--56 days after diagnosis vs. Treatment start within 28 days after diagnosis &
0.95 (0.89, 1.00); 1.10 &
1.34 &
1.45 \\
\addlinespace
Zhong \emph{et al.}\ (2024) &
Locally advanced proximal gastric cancer patients; 3-year disease-free survival & 
Robotic total gastrectomy vs. Laparoscopic total gastrectomy &
0.69 (0.36, 1.329); 1.33 &
1.02 &
2.53 \\
\bottomrule
\end{tabular}
\end{sidewaystable}


\newpage
\bibliographystyle{agsm}
\bibliography{paperpile}

@ARTICLE{VanderWeele2017-jj,
  title     = "Sensitivity analysis in observational research: Introducing the
               {E}-value",
  author    = "VanderWeele, Tyler J and Ding, Peng",
  journal   = "Ann. Intern. Med.",
  publisher = "American College of Physicians",
  volume    =  167,
  number    =  4,
  pages     = "268--274",
  month     =  aug,
  year      =  2017,
  language  = "en"
}

@ARTICLE{Ding2016-uo,
  title     = "Sensitivity analysis without assumptions",
  author    = "Ding, Peng and VanderWeele, Tyler J",
  journal   = "Epidemiology",
  publisher = "Ovid Technologies (Wolters Kluwer Health)",
  volume    =  27,
  number    =  3,
  pages     = "368--377",
  month     =  may,
  year      =  2016,
  language  = "en"
}

@ARTICLE{Smeenk2025-nk,
  title     = "Pembrolizumab hybrid dosing is non-inferior to flat dosing in
               advanced non-small cell lung cancer: a real-world, retrospective
               bicenter cohort study",
  author    = "Smeenk, Michiel M and van der Noort, Vincent and Hendrikx, Jeroen
               M A and Abedian Kalkhoran, Hanieh and Smit, Egbert F and Theelen,
               Willemijn S M E",
  journal   = "J. Immunother. Cancer",
  publisher = "BMJ",
  volume    =  13,
  number    =  2,
  pages     = "e010065",
  month     =  feb,
  year      =  2025,
  language  = "en"
}

@ARTICLE{Durand2024-ta,
  title    = "Safety of kidney transplantation from donors with {HIV}",
  author   = "Durand, Christine M and Massie, Allan and Florman, Sander and
              Liang, Tao and Rana, Meenakshi M and Friedman-Moraco, Rachel and
              Gilbert, Alexander and Stock, Peter and Mehta, Sapna A and Mehta,
              Shikha and Stosor, Valentina and Pereira, Marcus R and Morris,
              Michele I and Hand, Jonathan and Aslam, Saima and Malinis, Maricar
              and Haidar, Ghady and Small, Catherine B and Santos, Carlos A Q
              and Schaenman, Joanna and Baddley, John and Wojciechowski, David
              and Blumberg, Emily A and Ranganna, Karthik and Adebiyi,
              Oluwafisayo and Elias, Nahel and Castillo-Lugo, Jose A and
              Giorgakis, Emmanouil and Apewokin, Senu and Brown, Diane and
              Ostrander, Darin and Eby, Yolanda and Desai, Niraj and Naqvi,
              Fizza and Bagnasco, Serena and Watson, Natasha and Brittain, Erica
              and Odim, Jonah and Redd, Andrew D and Tobian, Aaron A R and
              Segev, Dorry L and {HOPE in Action Investigators}",
  journal  = "N. Engl. J. Med.",
  volume   =  391,
  number   =  15,
  pages    = "1390--1401",
  month    =  oct,
  year     =  2024,
  language = "en"
}

@ARTICLE{Rydbeck2023-ao,
  title     = "Survival in relation to time to start of curative treatment of
               colon cancer: A national register-based observational
               noninferiority study",
  author    = "Rydbeck, Daniel and Bock, David and Haglind, Eva and Angenete,
               Eva and Onerup, Aron",
  journal   = "Colorectal Dis.",
  publisher = "Wiley",
  volume    =  25,
  number    =  8,
  pages     = "1613--1621",
  month     =  aug,
  year      =  2023,
  language  = "en"
}

@ARTICLE{Althunian2017-dv,
  title     = "Defining the noninferiority margin and analysing noninferiority:
               An overview",
  author    = "Althunian, Turki A and de Boer, Anthonius and Groenwold, Rolf H H
               and Klungel, Olaf H",
  journal   = "Br. J. Clin. Pharmacol.",
  publisher = "John Wiley \& Sons, Ltd",
  volume    =  83,
  number    =  8,
  pages     = "1636--1642",
  month     =  aug,
  year      =  2017,
  language  = "en"
}

@ARTICLE{Lash2021-ri,
  title     = "Lash et al. Respond to ``Better Bias Analysis'' and ``Toward
               Better Bias Analysis''",
  author    = "Lash, Timothy L and Ahern, Thomas P and Collin, Lindsay J and
               Fox, Matthew P and MacLehose, Richard F",
  journal   = "Am. J. Epidemiol.",
  publisher = "Oxford University Press (OUP)",
  volume    =  190,
  number    =  8,
  pages     = "1622--1624",
  month     =  aug,
  year      =  2021,
  language  = "en"
}

@ARTICLE{Zhong2024-ty,
  title     = "Long-term survival outcomes of robotic total gastrectomy for
               locally advanced proximal gastric cancer: a prospective study",
  author    = "Zhong, Qing and Tang, Yi-Hui and Liu, Zhi-Yu and Zhang, Zhi-Quan
               and He, Qi-Chen and Li, Ping and Xie, Jian-Wei and Wang, Jia-Bin
               and Lin, Jian-Xian and Lu, Jun and Chen, Qi-Yue and Zheng,
               Chao-Hui and Huang, Chang-Ming",
  journal   = "Int. J. Surg.",
  publisher = "Ovid Technologies (Wolters Kluwer Health)",
  volume    =  110,
  number    =  7,
  pages     = "4132--4142",
  month     =  jul,
  year      =  2024,
  language  = "en"
}

@ARTICLE{Poole2020-qk,
  title     = "Commentary: Continuing the {E}-value's post-publication peer
               review",
  author    = "Poole, Charles",
  journal   = "Int. J. Epidemiol.",
  publisher = "Oxford University Press (OUP)",
  volume    =  49,
  number    =  5,
  pages     = "1497--1500",
  month     =  oct,
  year      =  2020,
  language  = "en"
}

@ARTICLE{Blum2020-qp,
  title     = "Use of {E}-values for addressing confounding in observational
               studies-an empirical assessment of the literature",
  author    = "Blum, Manuel R and Tan, Yuan Jin and Ioannidis, John P A",
  journal   = "Int. J. Epidemiol.",
  publisher = "Oxford University Press (OUP)",
  volume    =  49,
  number    =  5,
  pages     = "1482--1494",
  month     =  oct,
  year      =  2020,
  language  = "en"
}

@ARTICLE{Mauri2017-jl,
  title     = "Challenges in the design and interpretation of noninferiority
               trials",
  author    = "Mauri, Laura and D'Agostino, Sr, Ralph B",
  journal   = "N. Engl. J. Med.",
  publisher = "Massachusetts Medical Society",
  volume    =  377,
  number    =  14,
  pages     = "1357--1367",
  month     =  oct,
  year      =  2017,
  language  = "en"
}

@ARTICLE{Trinquart2019-hx,
  title     = "Applying the {E} value to assess the robustness of epidemiologic
               fields of inquiry to unmeasured confounding",
  author    = "Trinquart, Ludovic and Erlinger, Adrienne L and Petersen, Julie M
               and Fox, Matthew and Galea, Sandro",
  journal   = "Am. J. Epidemiol.",
  publisher = "Oxford University Press (OUP)",
  volume    =  188,
  number    =  6,
  pages     = "1174--1180",
  month     =  jun,
  year      =  2019,
  language  = "en"
}

@ARTICLE{Fox2020-vp,
  title     = "Commentary: The value of {E}-values and why they are not enough",
  author    = "Fox, Matthew P and Arah, Onyebuchi A and Stuart, Elizabeth A",
  journal   = "Int. J. Epidemiol.",
  publisher = "Oxford University Press (OUP)",
  volume    =  49,
  number    =  5,
  pages     = "1505--1506",
  month     =  oct,
  year      =  2020,
  language  = "en"
}

@ARTICLE{Ioannidis2019-fv,
  title     = "Limitations and misinterpretations of {E}-values for sensitivity
               analyses of observational studies",
  author    = "Ioannidis, John P A and Tan, Yuan Jin and Blum, Manuel R",
  journal   = "Ann. Intern. Med.",
  publisher = "American College of Physicians",
  volume    =  170,
  number    =  2,
  pages     = "108--111",
  month     =  jan,
  year      =  2019,
  language  = "en"
}

@ARTICLE{Greenland2020-yd,
  title     = "Commentary: An argument against {E}-values for assessing the
               plausibility that an association could be explained away by
               residual confounding",
  author    = "Greenland, Sander",
  journal   = "Int. J. Epidemiol.",
  publisher = "Oxford University Press (OUP)",
  volume    =  49,
  number    =  5,
  pages     = "1501--1503",
  month     =  oct,
  year      =  2020,
  language  = "en"
}

@ARTICLE{Westreich2017-rp,
  title     = "Transportability of trial results using inverse odds of sampling
               weights",
  author    = "Westreich, Daniel and Edwards, Jessie K and Lesko, Catherine R
               and Stuart, Elizabeth and Cole, Stephen R",
  journal   = "Am. J. Epidemiol.",
  publisher = "Am J Epidemiol",
  volume    =  186,
  number    =  8,
  pages     = "1010--1014",
  month     =  oct,
  year      =  2017,
  language  = "en"
}

@ARTICLE{Lipsitch2010-or,
  title     = "Negative controls: a tool for detecting confounding and bias in
               observational studies",
  author    = "Lipsitch, Marc and Tchetgen Tchetgen, Eric and Cohen, Ted",
  journal   = "Epidemiology",
  publisher = "Epidemiology",
  volume    =  21,
  number    =  3,
  pages     = "383--388",
  month     =  may,
  year      =  2010,
  language  = "en"
}

\newpage
\section*{Figures}

\end{document}